\documentclass[a4paper]{jpconf}
\usepackage{graphicx}
\begin{document}
\title{A Data-driven dE/dx Simulation with Normalizing Flow}

\author{Wenxing Fang$^{1*}$, Weidong Li$^1$, Xiaobin Ji$^1$, Shengsen Sun$^1$, Tong Chen$^1$, Fang Liu$^1$, Xiaoling Li$^2$, Kai Zhu$^1$, Tao Lin$^1$,and Jinfa Qiu$^1$ }

\address{$^1$ Institute of High Energy Physics, Beijing 100049, People’s Republic of China}
\address{$^2$ Institute of Frontier and Interdisciplinary Science, Shandong University, Qingdao, Shandong, China}
\ead{fangwx@ihep.ac.cn}

\begin{abstract}
In high-energy physics, precise measurements rely on highly reliable detector simulations. Traditionally, these simulations involve incorporating experiment data to model detector responses and fine-tuning them. However, due to the complexity of the experiment data, tuning the simulation can be challenging. One crucial aspect for charged particle identification is the measurement of energy deposition per unit length (referred to as dE/dx). This paper proposes a data-driven dE/dx simulation method using the Normalizing Flow technique, which can learn the dE/dx distribution directly from experiment data. By employing this method, not only can the need for manual tuning of the dE/dx simulation be eliminated, but also high-precision simulation can be achieved.

\end{abstract}

\section{Introduction}

The measurement of dE/dx in gas detectors, such as drift chambers or time projection chambers (TPCs), plays a crucial role in charged particle identification (PID). However, accurately simulating the energy loss of charged particles in thin gas remains a challenge for Geant4\cite{Geant4}. As a result, the conventional approach for dE/dx simulation relies on histogram sampling\cite{Cao_2010}. This method involves categorizing experimental dE/dx measurements into different histograms based on particle properties. For instance, the $\beta\gamma$ range is divided into multiple regions, with each region having its own histogram of dE/dx distribution. During simulation, dE/dx values are sampled from the appropriate histograms, depending on the particle's properties. While this approach achieves reasonably precise dE/dx simulation, it requires non-trivial fine-tuning, including optimizing the regions for each property.

To address the challenges associated with fine-tuning in dE/dx simulation, this study proposes a data-driven approach using deep learning techniques. Specifically, we employ the Normalizing Flow\cite{NF} technique, which has been extensively utilized in simulation tasks such as fast calorimeter simulation\cite{CaloFlow}. In this study, we utilize experimental data from The Beijing Spectrometer III (BESIII)\cite{BES3}, a precision measurement experiment focused on particle physics in the tau-charm region, to perform dE/dx simulation using the Normalizing Flow technique.

\section{Data sample}
\label{dataSample}
To conduct our study, we utilized $e^{+}e^{-}\rightarrow J/\psi$ data obtained from the BESIII experiment in 2018. The data underwent reconstruction using the BESIII offline software (BOSS)\cite{BOSS}. In order to obtain pure hadron samples, we focused on three specific physics processes: the pion sample derived from $J/\psi \rightarrow \rho\pi \rightarrow \pi\pi\pi$, the kaon sample obtained from $J/\psi \rightarrow K_{s}^{0}K\pi \rightarrow K \pi \pi \pi$, and the proton(anti) sample extracted from $J/\psi \rightarrow pp\pi\pi$.

Considering that this study represents our initial attempt at simulating dE/dx, we focused on a track-level simulation. This choice was driven by two factors: the relative ease of implementing a track-level simulation compared to a hit-level simulation, and the fact that track-level simulation satisfies the requirements for subsequent physics analysis. For each type of particle, we saved the following information for training purposes: reconstructed momentum, polar angle ($\theta$), dE/dx values, and the number of hits (nHits) utilized for dE/dx reconstruction.

To ensure the quality and reliability of our training data, we applied smoothing techniques to the event numbers in both the momentum and $\theta$ dimensions. Table \ref{tab_data} provides an overview of the number of events included in the training and testing datasets.
\begin{table}[h!]
\centering
\begin{tabular}{ |c|c|c|c|c|c|c| } 
 \hline
               & $\pi{+}$ & $\pi{-}$ & $K^{+}$ & $K^{-}$ & $p^{+}$ & $p^{-}$ \\ \hline 
 Training data & 1M       & 1M       & 0.5M    & 0.5M    & 2M      & 2M      \\ \hline
 Testing data  & 0.4M     & 0.4M     & 0.2M    & 0.2M    & 0.9M    & 0.9M    \\ 
 \hline
\end{tabular}
\caption{The number of events for training and testing dataset in million (M).}
\label{tab_data}
\end{table}

\section{Simulation model}
\label{Model}
\begin{figure}
\begin{center}
\includegraphics[width=0.9\textwidth]{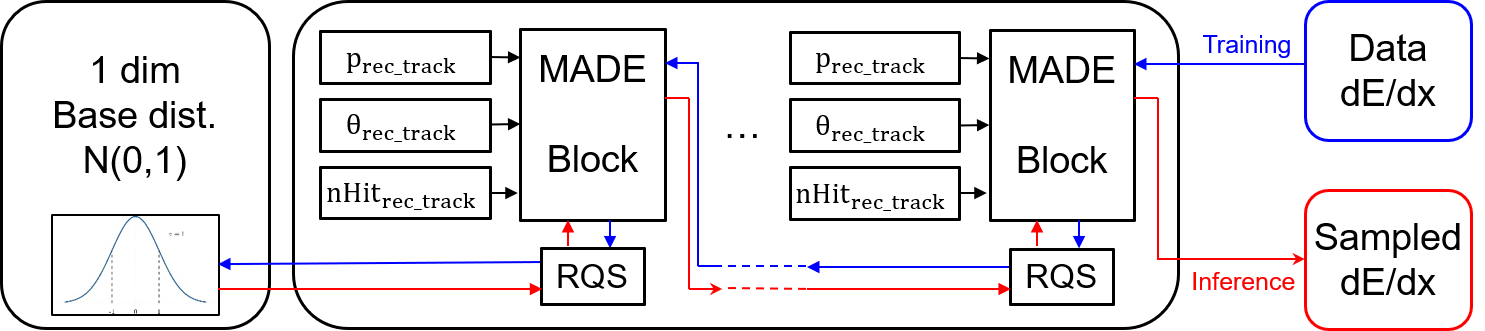}
\caption{Schematic view of the Normalizing Flow model. The training process follows blue arrows, inference process (or simulation) follows red arrows.} \label{model_NF}
\end{center}
\end{figure}

In this study, we employed six Normalizing Flow-based neural networks to simulate track-level dE/dx for $\pi^{+}, \pi^{-}, K^{+}, K^{-}, p^{+},$ and $p^{-}$ particles, respectively. These six models share the same neural network architecture, as depicted in Figure \ref{model_NF}.

The underlying concept of the model is to establish a bijective transformation between the conditional dE/dx distribution and a base distribution, which, in our case, is the standard normal distribution. The core of the model consists of transformation functions based on the Rational Quadratic Spline (RQS)\cite{RQS}, along with the MADE\cite{MADE} neural network blocks responsible for learning the parameters of the RQS transformations. The model configuration involves employing six identical MADE blocks consecutively. Each MADE block comprises an input layer with 64 neurons, three hidden layers with 64 neurons each, and an output layer with 23 neurons, determined by the eight bins of the RQS transformation.

During the training stage, the dE/dx values, together with the conditional data encompassing momentum, $\theta$, and nHits, are inputted and subsequently transformed to achieve a final standard normal distribution. For the inference stage, the process is reversed, where random numbers are sampled from the standard normal distribution and transformed back to yield simulated dE/dx values based on the given conditions of momentum, $\theta$, and nHits. We trained the models for 100 epochs using a batch size of 128 and the Adam optimizer with a learning rate of $10^{-3}$. The learning rate was halved every 10 epochs.

\section{Performance}
\label{MLSim_results}
\begin{figure}
\begin{center}
\includegraphics[width=0.4\textwidth]{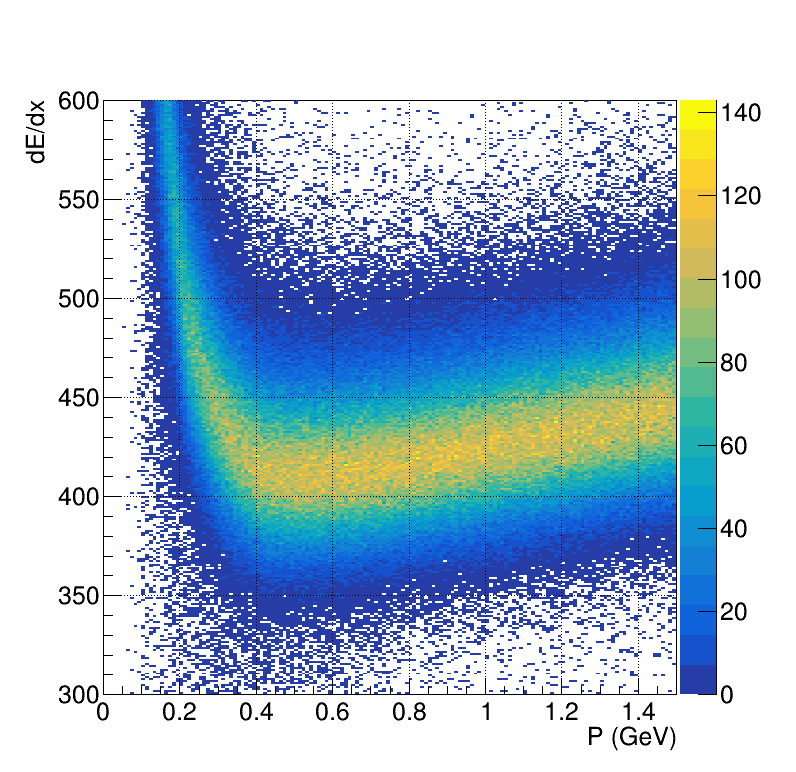}
\includegraphics[width=0.4\textwidth]{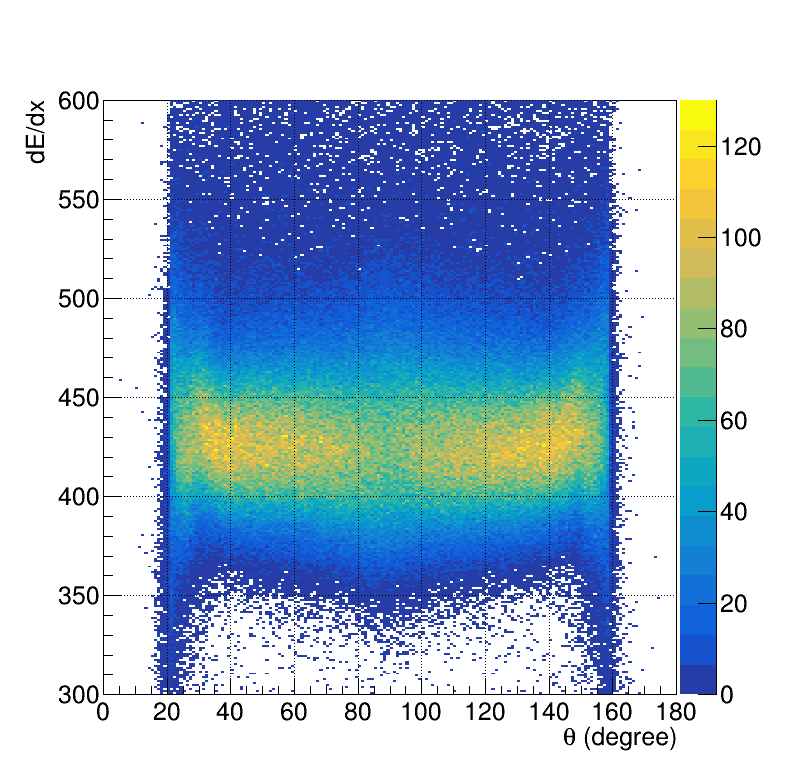}
\includegraphics[width=0.4\textwidth]{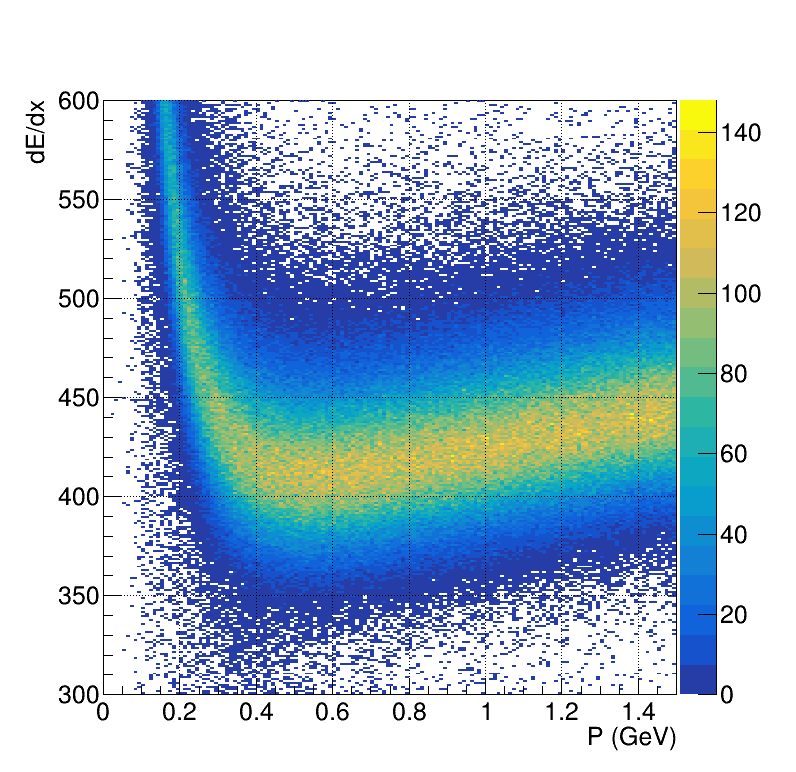}
\includegraphics[width=0.4\textwidth]{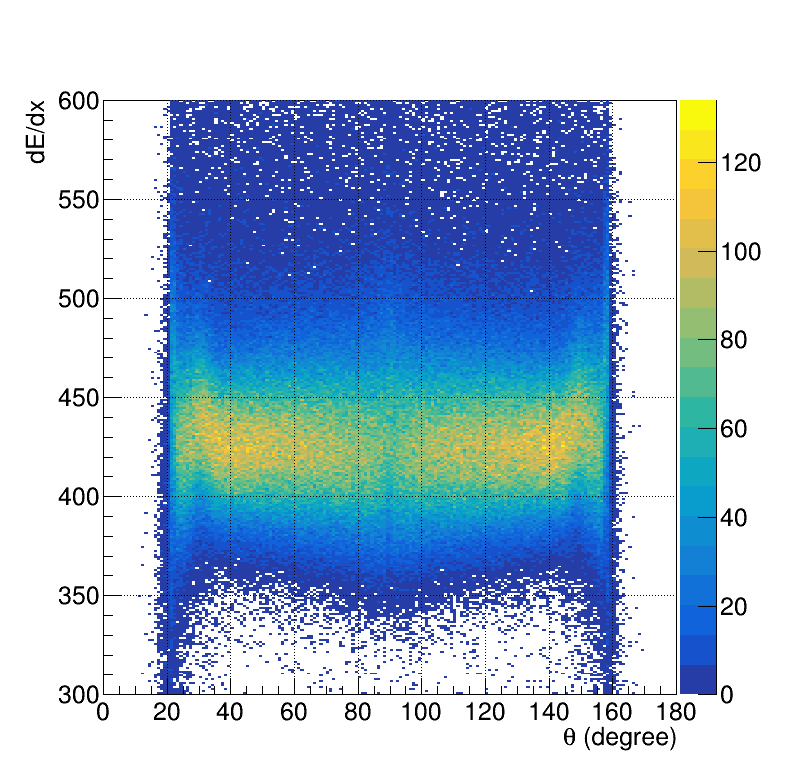}
\caption{The dE/dx distribution of $\pi^{+}$. The left (right) plots are dE/dx versus momentum ($\mathrm{\theta}$). The top (bottom) plots for simulated (data).} \label{fig_pip_dEdx_compare}
\end{center}
\end{figure}

We initiated our analysis by examining the simulated dE/dx distribution for each particle species and observed remarkable similarity to the actual data. Figure \ref{fig_pip_dEdx_compare} illustrates an example from $\pi^{+}$, demonstrating the dE/dx distribution in comparison to momentum and $\theta$ for both simulated and real data. The plots exhibit minimal discernible differences between the simulated and real data, indicating a high degree of fidelity in our simulation. Similar plots for $K^{+}$ and $p^{+}$ can be found in Appendix \ref{app_dEdx}.

To quantitatively assess the quality of the simulation, we employed a widely used metric in physics analysis: the dE/dx particle identification (PID) efficiency. To calculate the dE/dx PID efficiency, we compute $\chi_{dE/dx}$ (Equation \ref{eq_chi_dEdx}) for various particle hypotheses. Here, $dE/dx_{measured}$ represents the measured dE/dx value, while $dE/dx_{exp}^{i}$ and $\sigma_{dE/dx}^{i}$ denote the expected dE/dx and dE/dx resolution for different particle hypotheses. For the purpose of this evaluation, we focus on the pion, kaon, and proton(anti) hypotheses, as these are typically sufficient for most physics analyses.

For instance, when evaluating the dE/dx PID for a pion, it should exhibit the minimum value of $|\chi_{dE/dx}^{\pi}|$ among various hypotheses, and this value should be less than 4. 

\begin{equation}
\label{eq_chi_dEdx}
\chi_{dE/dx}^{i}=\frac{{dE/dx}_{measured}-{dE/dx}_{exp}^{i}}{\sigma_{dE/dx}^{i}}
\end{equation}

\begin{figure}
\begin{center}
\includegraphics[width=0.4\textwidth]{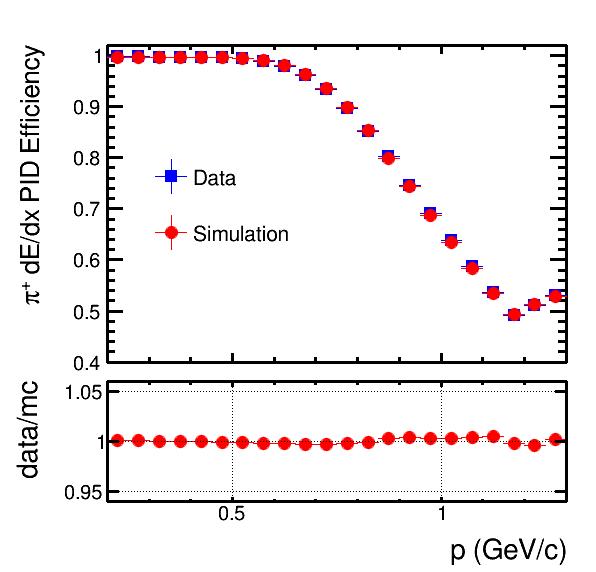}
\includegraphics[width=0.4\textwidth]{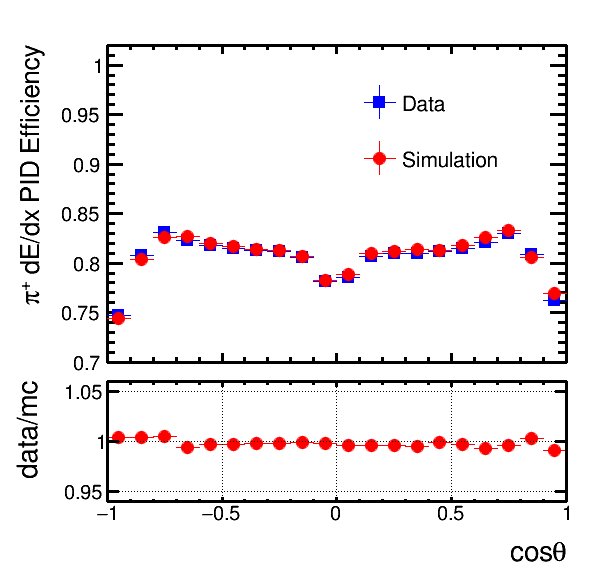}
\caption{The dE/dx PID efficiency of $\pi^{+}$ versus momentum (cos$\theta$) in the left (right) plot. The blue squares are for data, the red circles are for simulation.} 
\label{fig_pip_pid}
\end{center}
\end{figure}

To compare the dE/dx PID efficiency between the data and the simulation, we generated corresponding Monte Carlo samples using BOSS. The dE/dx values for these Monte Carlo samples were simulated using the neural network. Figure \ref{fig_pip_pid} presents the dE/dx PID efficiency for $\pi^{+}$ and showcases a comprehensive comparison between the data and the simulation. Notably, the efficiency is evaluated across both the momentum and cos$\theta$ dimensions.
The plot demonstrates a remarkable consistency in the efficiency between the data and the simulation. This agreement validates the high fidelity of the simulation, as it accurately reproduces the dE/dx PID efficiency observed in the actual data.

\begin{figure}
\begin{center}
\includegraphics[width=0.4\textwidth]{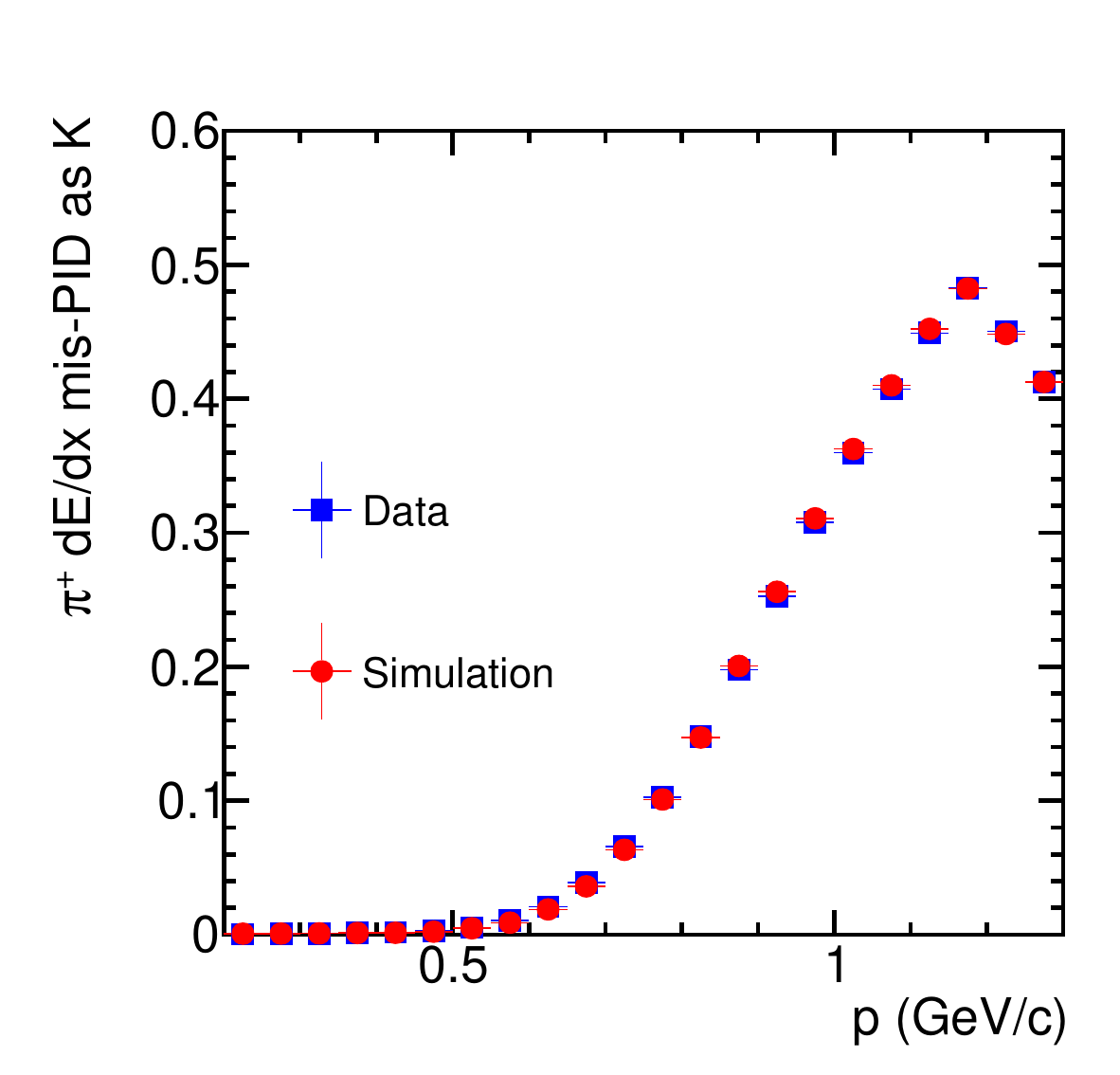}
\includegraphics[width=0.4\textwidth]{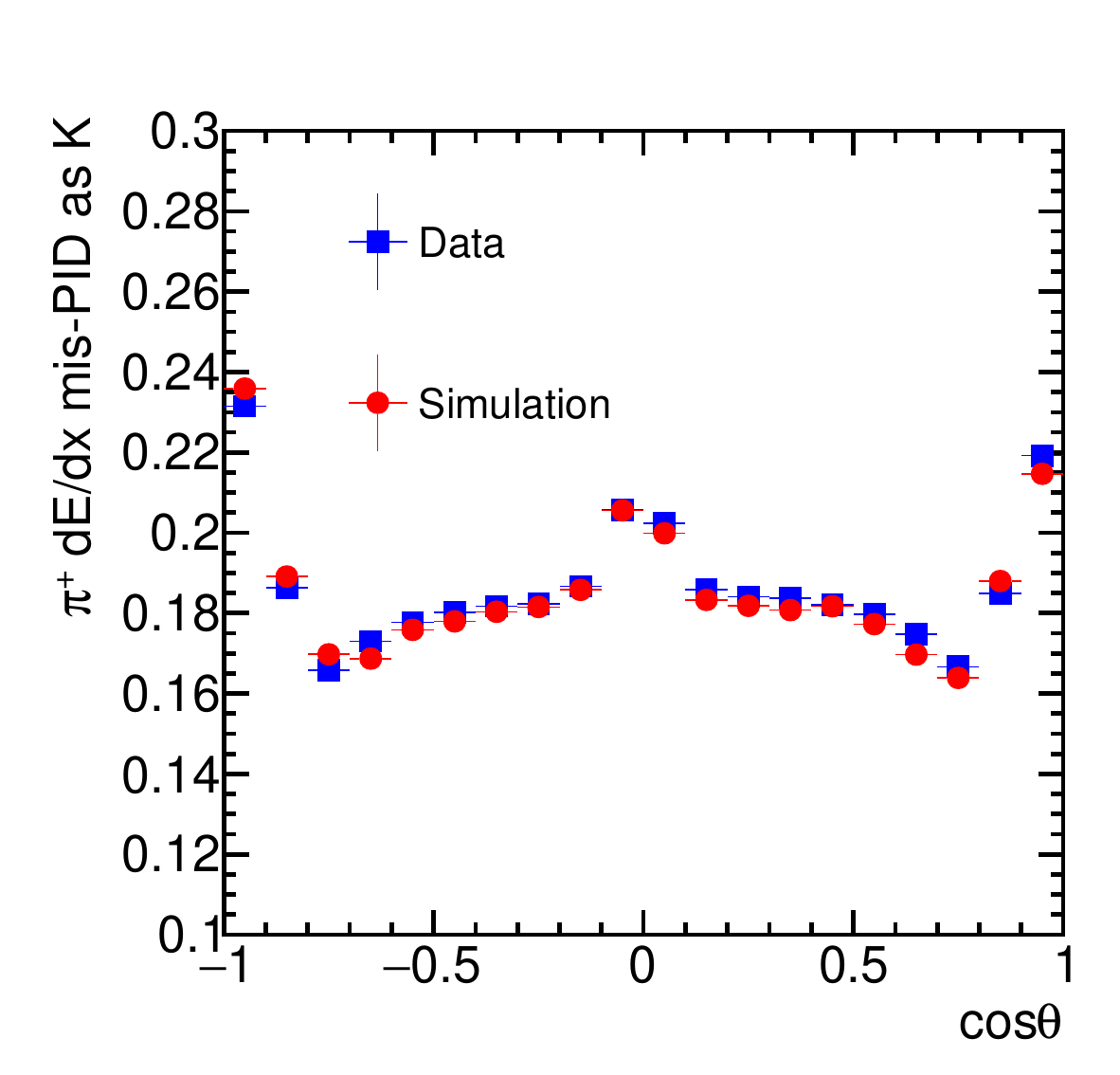}
\caption{The efficiency of $\pi^{+}$ misidentified as $K^{+}$ by using dE/dx. The left (right) is the miss identified efficiency versus momentum (cos$\theta$). The blue squares are for data, the red circles are for simulation.} \label{fig_pip_misPID_K}
\end{center}
\end{figure}

In physics analysis, not only the dE/dx PID efficiency but also the dE/dx misidentified PID efficiency plays a crucial role. It provides insights into the misidentification of particles based on their dE/dx values. Specifically, we can examine the efficiency of $\pi^{+}$ being misidentified as $K^{+}$, which is shown in Figure \ref{fig_pip_misPID_K}. This figure presents a comparison between the data and the simulation, highlighting the misidentified efficiency across both momentum and cos$\theta$ dimensions. Notably, the misidentified efficiency is consistent between the data and the simulation, reinforcing the fidelity of the simulation in capturing the misidentification behavior.

\begin{figure}
\begin{center}
\includegraphics[width=0.4\textwidth]{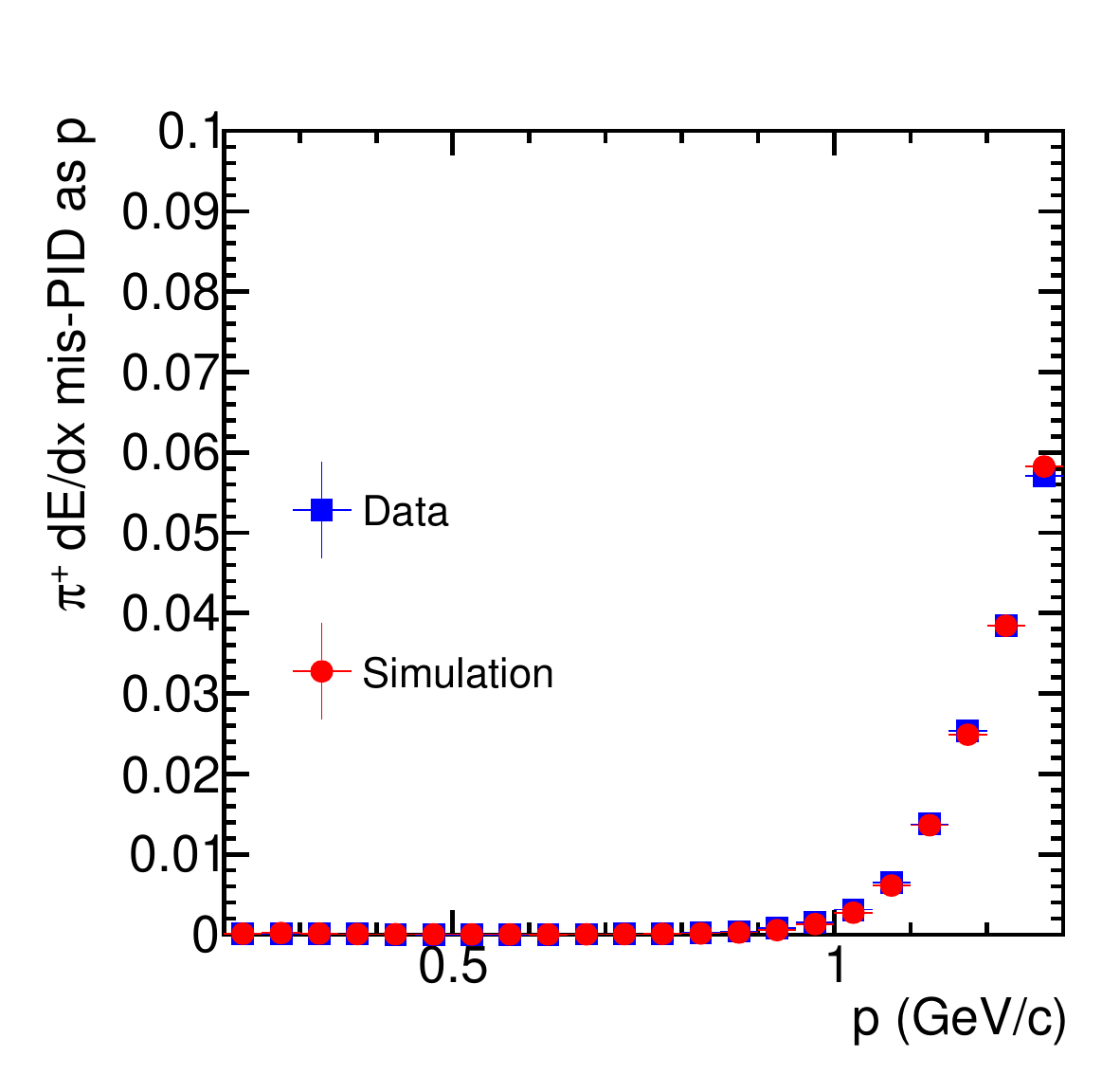}
\includegraphics[width=0.4\textwidth]{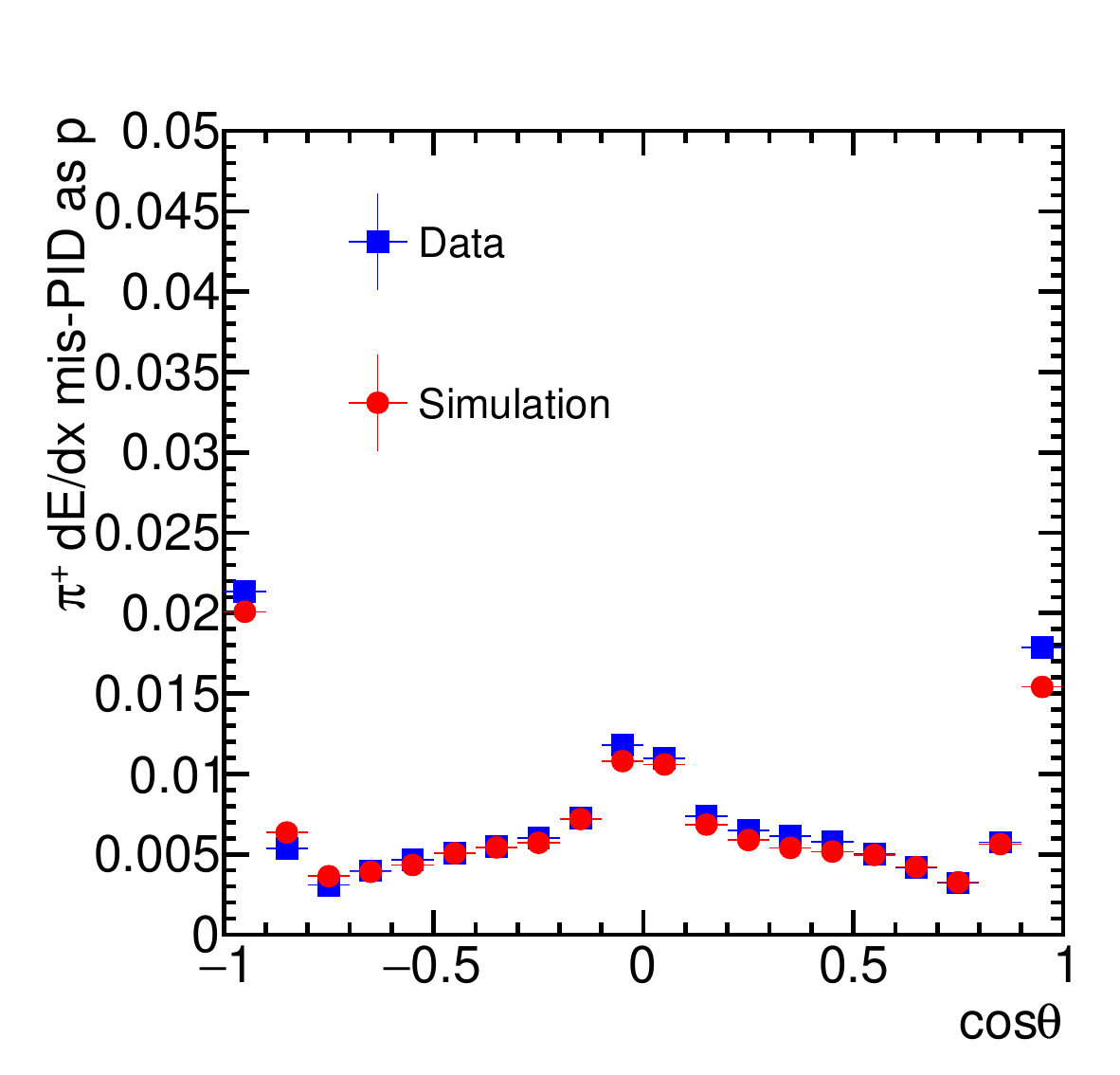}
\caption{The efficiency of $\pi^{+}$ misidentified as $p^{+}$ by using dE/dx. The left (right) is the miss identified efficiency versus momentum (cos$\theta$). The blue squares are for data, the red circles are for simulation.} 
\label{fig_pip_misPID_P}
\end{center}
\end{figure}

Furthermore, Figure \ref{fig_pip_misPID_P} showcases a similar analysis, focusing on the efficiency of $\pi^{+}$ being misidentified as $p^{+}$. Once again, we observe a consistent trend between the data and the simulation, demonstrating the agreement in the misidentified efficiency in both momentum and cos$\theta$ dimensions.
These results further validate the accuracy of the simulation in reproducing the misidentified PID efficiencies observed in the actual data.

In principle, the dE/dx is primarily dependent on the characteristics of the particle itself. Therefore, a trained dE/dx simulation neural network, such as the one trained using $J/\psi \rightarrow \rho\pi \rightarrow \pi\pi\pi$ data, is general and can be applied to simulate the dE/dx of pions in other physics processes.

\begin{figure}
\begin{center}
\includegraphics[width=0.4\textwidth]{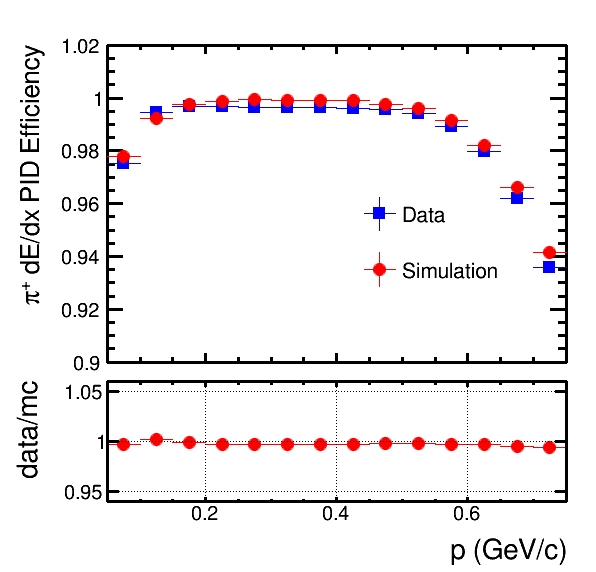}
\includegraphics[width=0.4\textwidth]{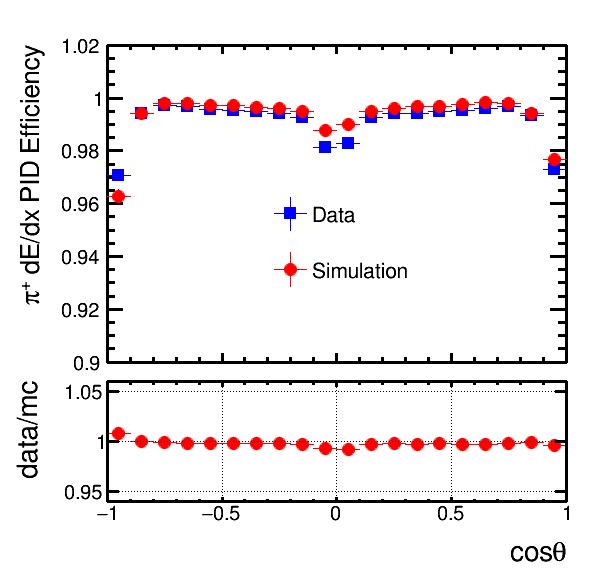}
\caption{The dE/dx PID efficiency of $\pi^{+}$ versus momentum (cos$\theta$) in the left (right) plot. The blue squares are for data, the red circles are for simulation.} 
\label{fig_pip_pid_pppipi}
\end{center}
\end{figure}

To verify the reliability of the trained neural network, we can perform a safety check by examining the dE/dx PID efficiency of $\pi^{+}$ from the $J/\psi \rightarrow pp\pi\pi$ process. In this case, the dE/dx simulation for pions is based on the model trained with $J/\psi \rightarrow \rho\pi \rightarrow \pi\pi\pi$ data. Figure \ref{fig_pip_pid_pppipi} presents the comparison between the data and the simulation in terms of the dE/dx PID efficiency for $\pi^{+}$.
As expected, the efficiency between the data and the simulation is in good agreement, further confirming the generalizability of the trained dE/dx simulation neural network. This consistency demonstrates that the neural network trained with $J/\psi \rightarrow \rho\pi \rightarrow \pi\pi\pi$ data can accurately simulate the dE/dx of pions in different physics processes.

\begin{figure}
\begin{center}
\includegraphics[width=0.4\textwidth]{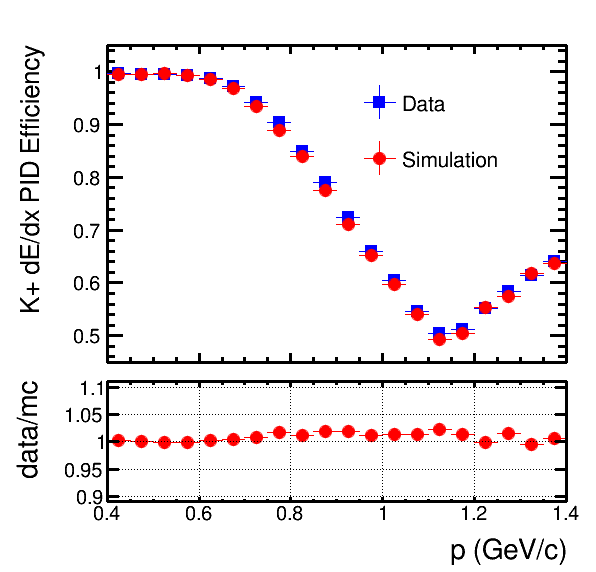}
\includegraphics[width=0.4\textwidth]{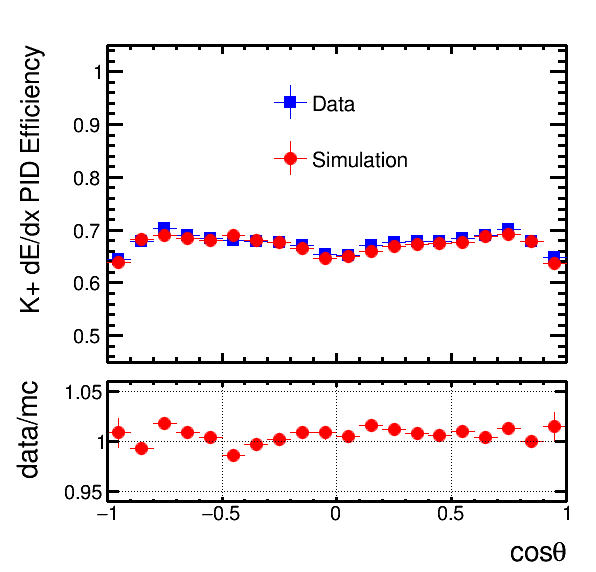}
\caption{The dE/dx PID efficiency of $K^{+}$ versus momentum (cos$\theta$) in the left (right) plot. The blue squares are for data, the red circles are for simulation.} 
\label{fig_kp_pid}
\end{center}
\end{figure}

\begin{figure}
\begin{center}
\includegraphics[width=0.4\textwidth]{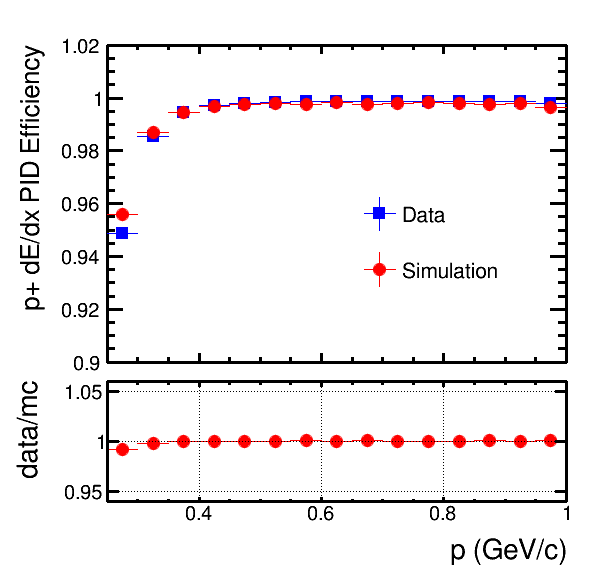}
\includegraphics[width=0.4\textwidth]{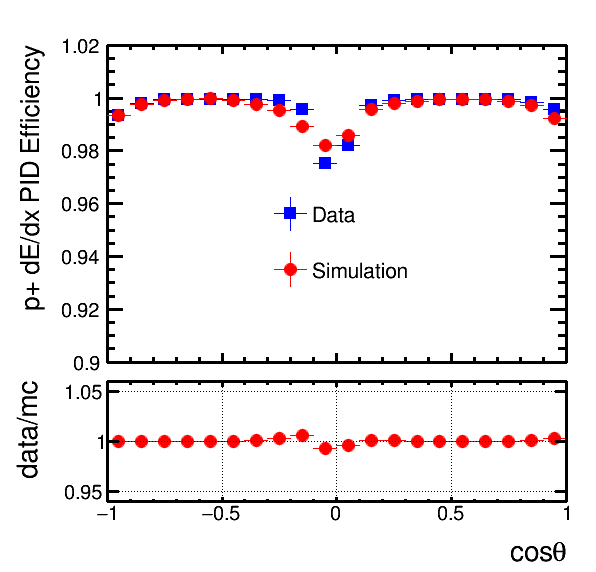}
\caption{The dE/dx PID efficiency of $p^{+}$ versus momentum (cos$\theta$) in the left (right) plot. The blue squares are for data, the red circles are for simulation.} 
\label{fig_pp_pid}
\end{center}
\end{figure}

In a similar manner, the performance of the dE/dx simulation for $K^{+}$ and $p^{+}$ can be examined. The results demonstrate a good agreement between the data and the simulation, further validating the accuracy of the simulation.
For instance, Figure \ref{fig_kp_pid} displays the dE/dx PID efficiency for $K^{+}$, while Figure \ref{fig_pp_pid} showcases the dE/dx PID efficiency for $p^{+}$. These figures provide a comparison between the data and the simulation, illustrating the consistency in the efficiency measurements.

Moreover, the dE/dx PID efficiency for $\pi^{-}$, $K^{-}$, and $p^{-}$ can be found in Appendix \ref{app_dEdx_pid}, providing a comprehensive analysis of the efficiency measurements for these particle species.

\section{Conclusion}
\label{Conclusion}

The dE/dx measurement is a crucial tool for particle identification (PID) and plays a significant role in precision measurement experiments. In order to achieve reliable and high-precision dE/dx simulation without the need for extensive fine-tuning, a data-driven approach utilizing the Normalizing Flow technique has been proposed. This study utilized experimental data from the BESIII experiment, focusing on the identification of $\pi^{\pm}$, $K^{\pm}$, and $p^{\pm}$.
The results of this study demonstrate the exceptional fidelity of the simulated dE/dx distribution. The simulated dE/dx accurately reproduces the experimental data, indicating the high-quality performance of the proposed data-driven method.
To further evaluate the method's suitability for physics analysis, a comparison of the dE/dx PID efficiency between the data and the simulation was conducted. The overall agreement and consistency between the data and the simulation provide strong evidence that our method is well-suited for high-precision dE/dx simulation.
We firmly believe that our data-driven approach, combined with the Normalizing Flow technique, will be highly advantageous for experiments utilizing drift chambers or Time Projection Chambers (TPCs) for dE/dx-based particle identification. By providing reliable and accurate dE/dx simulation, our method contributes to enhancing the precision and reliability of physics analyses in these experiments.


\ack
This work is supported by National Natural Science Foundation of China under Grant No.12205322, National Key R$\&$D Program of China under Contract No.2020YFA0406304, the Innovation Project of the Institute of High Energy Physics under Grant No.E15455U2.

\section*{References}
\providecommand{\newblock}{}

\clearpage
\section*{Appendix}
\subsection{dE/dx distribution}
\label{app_dEdx}

\begin{figure}[h!]
\begin{center}
\includegraphics[width=0.35\textwidth]{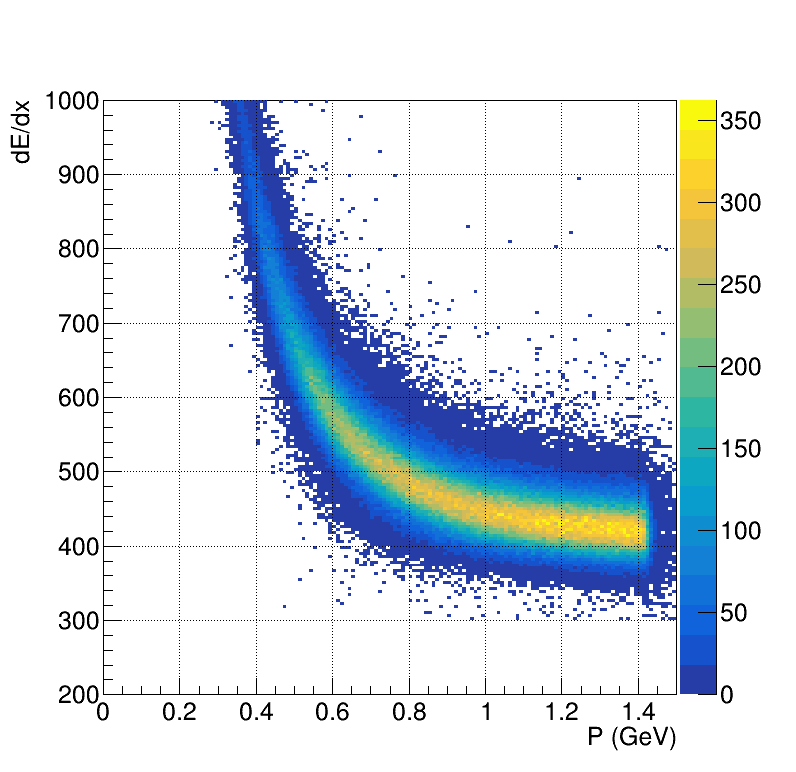}
\includegraphics[width=0.35\textwidth]{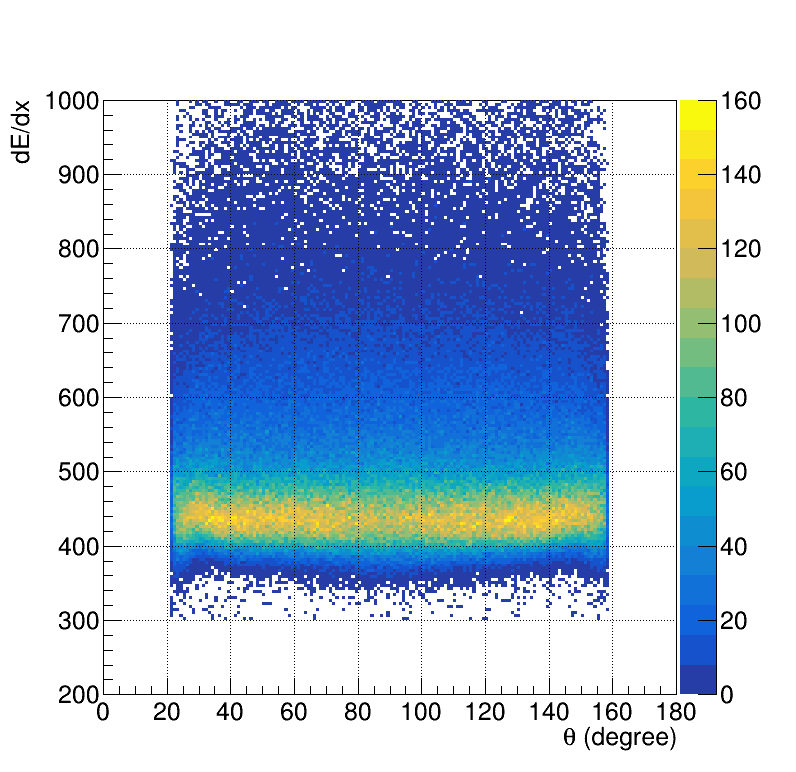}
\includegraphics[width=0.35\textwidth]{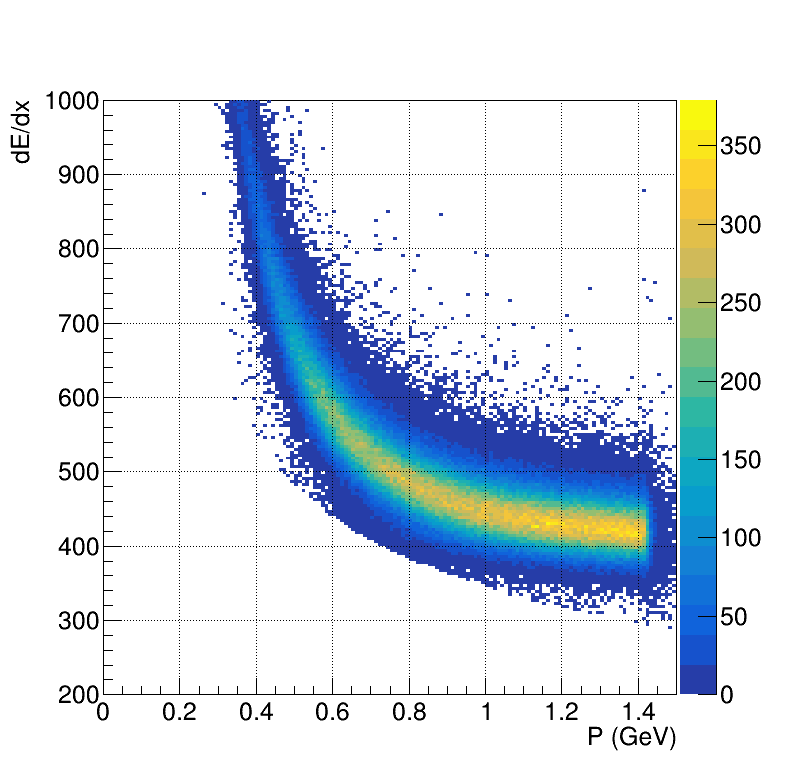}
\includegraphics[width=0.35\textwidth]{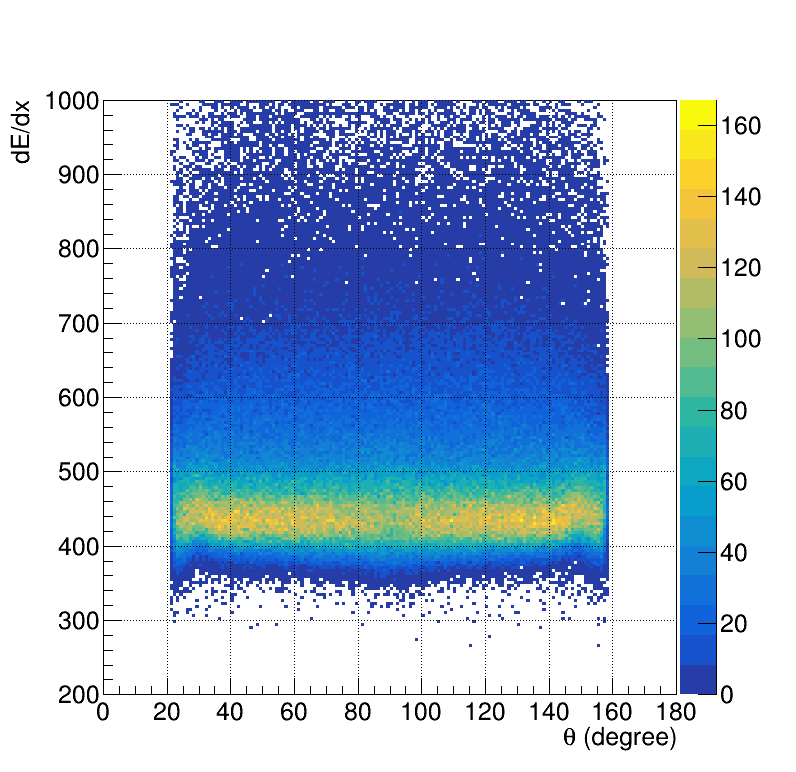}
\caption{The dE/dx distribution of $K^{+}$. The left (right) plots are dE/dx versus momentum ($\mathrm{\theta}$). The top (bottom) plots for simulated (data).} \label{fig_kp_dEdx_compare}
\end{center}
\end{figure}

\begin{figure}[h!]
\begin{center}
\includegraphics[width=0.35\textwidth]{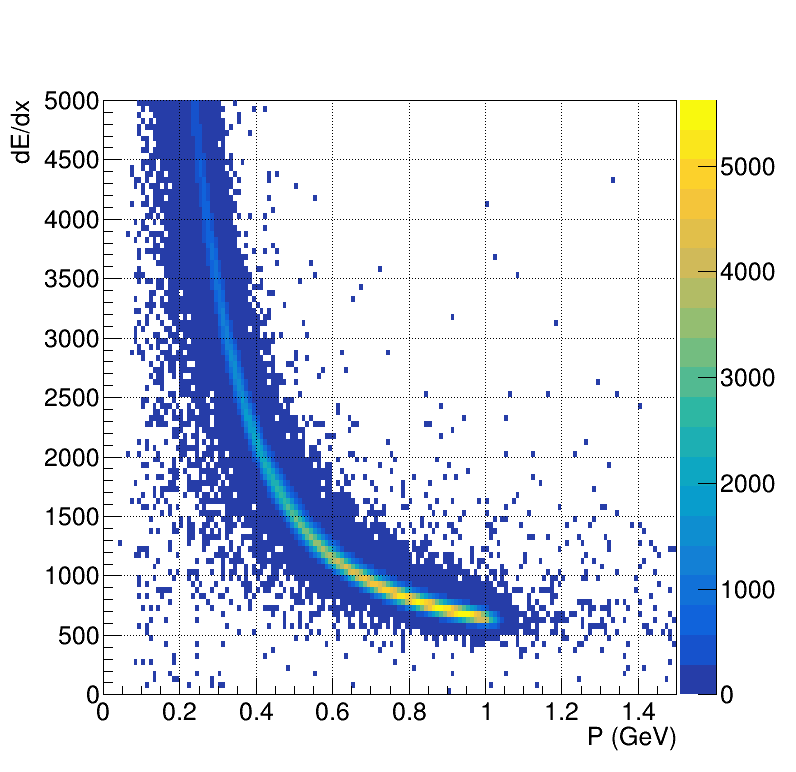}
\includegraphics[width=0.35\textwidth]{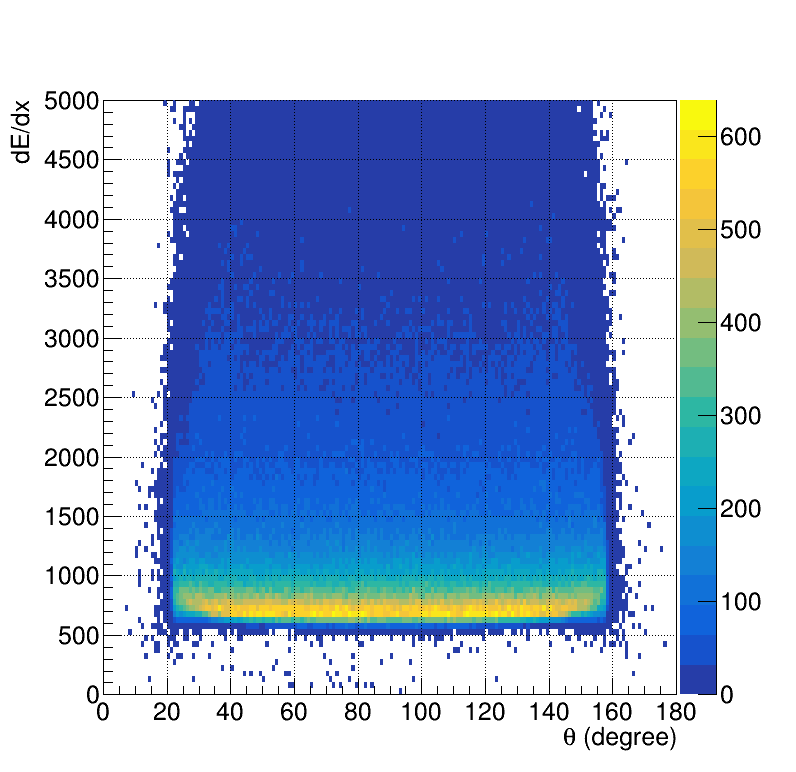}
\includegraphics[width=0.35\textwidth]{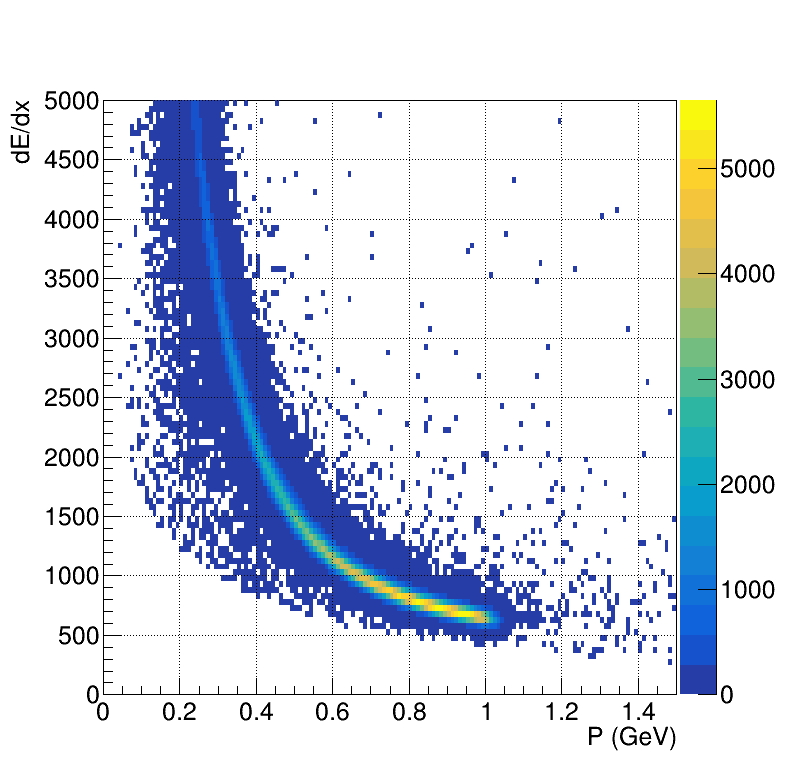}
\includegraphics[width=0.35\textwidth]{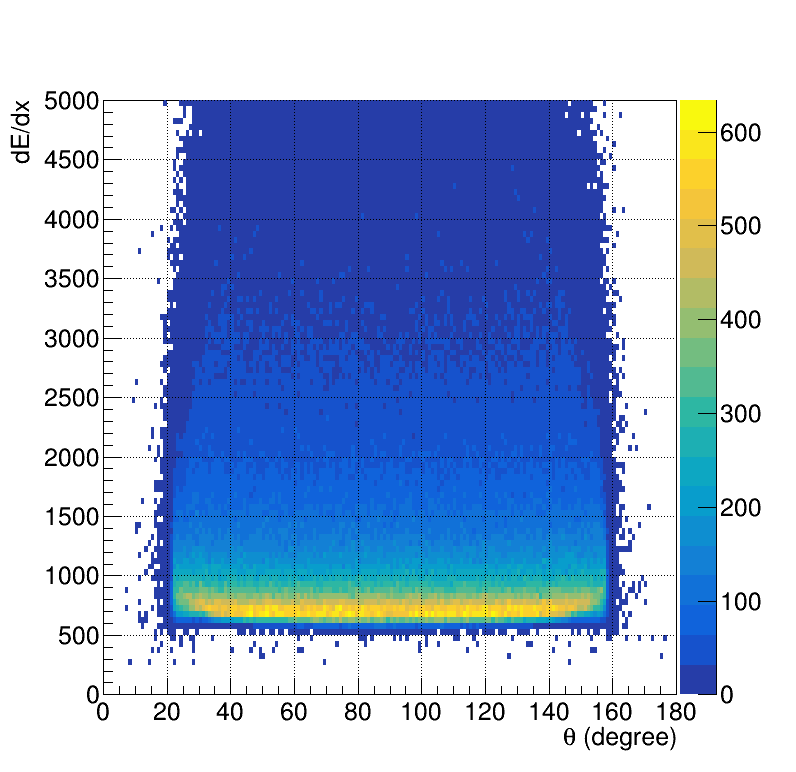}
\caption{The dE/dx distribution of $p^{+}$. The left (right) plots are dE/dx versus momentum ($\mathrm{\theta}$). The top (bottom) plots for simulated (data).} \label{fig_pp_dEdx_compare}
\end{center}
\end{figure}

\subsection{dE/dx PID efficiency}
\label{app_dEdx_pid}

\begin{figure}[h!]
\begin{center}
\includegraphics[width=0.4\textwidth]{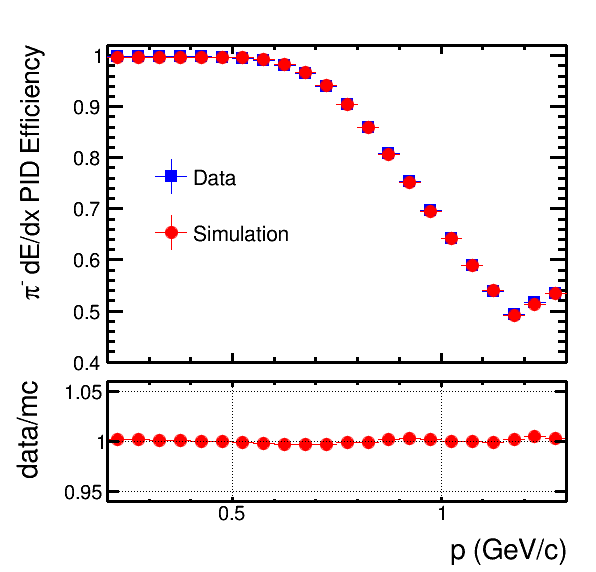}
\includegraphics[width=0.4\textwidth]{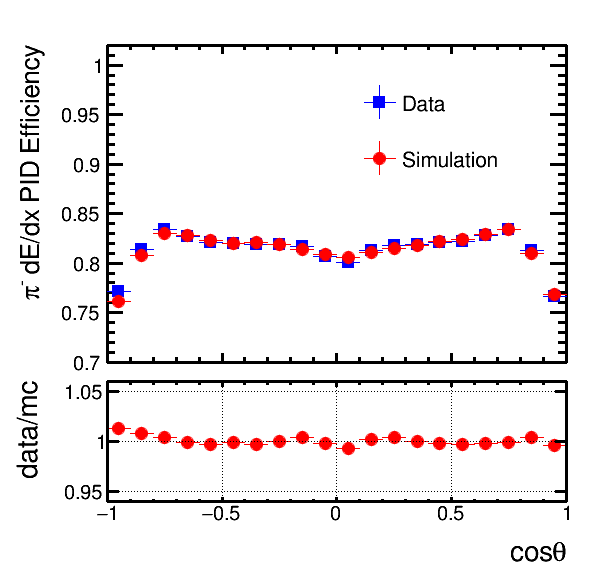}
\caption{The dE/dx PID efficiency of $\pi^{-}$ versus momentum (cos$\theta$) in the left (right) plot. The blue squares are for data, the red circles are for simulation.} 
\label{fig_pim_pid}
\end{center}
\end{figure}

\begin{figure}[h!]
\begin{center}
\includegraphics[width=0.4\textwidth]{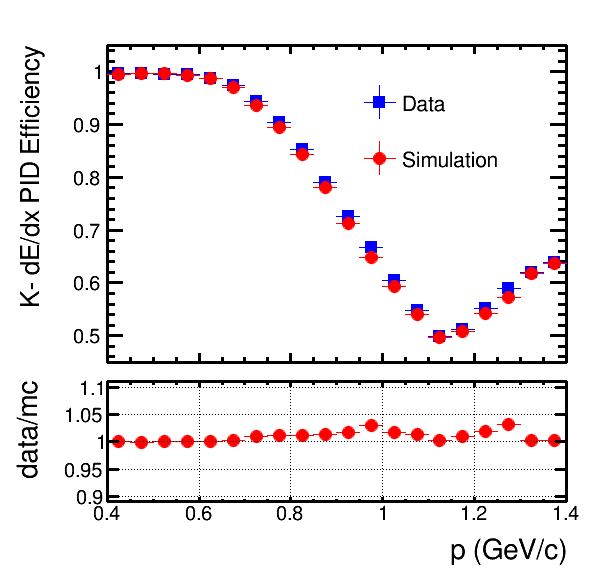}
\includegraphics[width=0.4\textwidth]{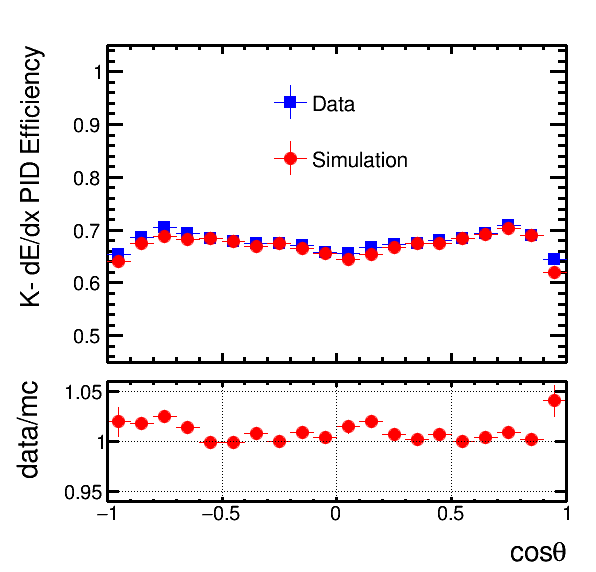}
\caption{The dE/dx PID efficiency of $K^{-}$ versus momentum (cos$\theta$) in the left (right) plot. The blue squares are for data, the red circles are for simulation.} 
\label{fig_km_pid}
\end{center}
\end{figure}

\begin{figure}[h!]
\begin{center}
\includegraphics[width=0.4\textwidth]{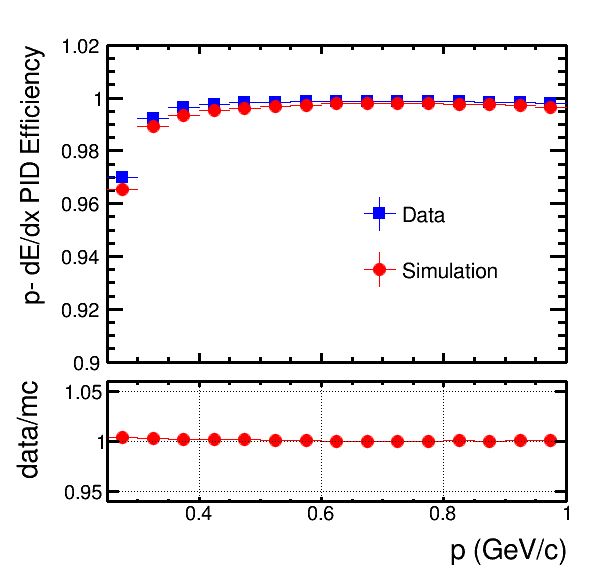}
\includegraphics[width=0.4\textwidth]{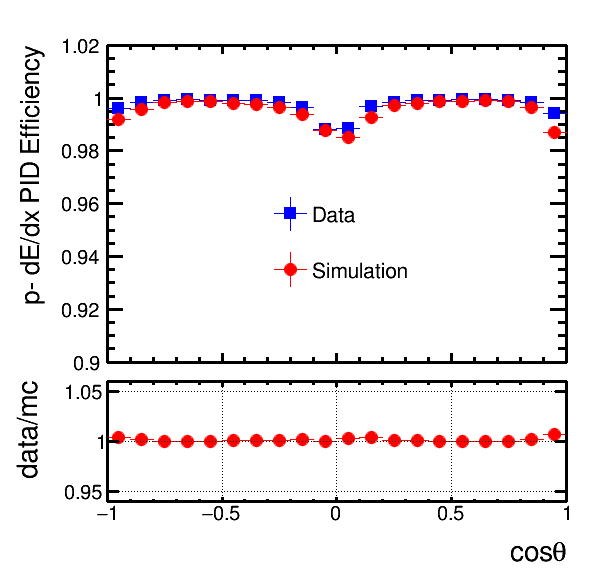}
\caption{The dE/dx PID efficiency of $p^{-}$ versus momentum (cos$\theta$) in the left (right) plot. The blue squares are for data, the red circles are for simulation.} 
\label{fig_pm_pid}
\end{center}
\end{figure}

\end{document}